\documentclass{article}
\usepackage{lscape,epsfig}
\usepackage{graphicx}
\usepackage{amssymb}
\usepackage{amsmath}
\usepackage{natbib}
\textheight=22cm \DeclareSymbolFont{ppa}{OT1}{ppl}{m}{it}
\DeclareMathSymbol{\vv}{\mathalpha}{ppa}{'166}

minus 2.3mu \thickmuskip = 2.6mu plus 2mu minus 2.6mu

\begin{document}

\newcommand{\dd}{\,{\rm d}}
\newcommand{\ie}{{\it i.e.},\,}
\newcommand{\etal}{{\it $et$ $al$.\ }}
\newcommand{\eg}{{\it e.g.},\,}
\newcommand{\cf}{{\it cf.\ }}
\newcommand{\vs}{{\it vs.\ }}
\newcommand{\zdot}{\makebox[0pt][l]{.}}
\newcommand{\up}[1]{\ifmmode^{\rm #1}\else$^{\rm #1}$\fi}
\newcommand{\dn}[1]{\ifmmode_{\rm #1}\else$_{\rm #1}$\fi}
\newcommand{\upd}{\up{d}}
\newcommand{\uph}{\up{h}}
\newcommand{\upm}{\up{m}}
\newcommand{\ups}{\up{s}}
\newcommand{\arcd}{\ifmmode^{\circ}\else$^{\circ}$\fi}
\newcommand{\arcm}{\ifmmode{'}\else$'$\fi}
\newcommand{\arcs}{\ifmmode{''}\else$''$\fi}
\newcommand{\MS}{{\rm M}\ifmmode_{\odot}\else$_{\odot}$\fi}
\newcommand{\RS}{{\rm R}\ifmmode_{\odot}\else$_{\odot}$\fi}
\newcommand{\LS}{{\rm L}\ifmmode_{\odot}\else$_{\odot}$\fi}

\newcommand{\Abstract}[2]{{\footnotesize\begin{center}ABSTRACT\end{center}
\vspace{1mm}\par#1\par \noindent {~}{\it #2}}}

\newcommand{\TabCap}[2]{\begin{center}\parbox[t]{#1}{\begin{center}
  \small {\spaceskip 2pt plus 1pt minus 1pt T a b l e}
  \refstepcounter{table}\thetable \\[2mm]
  \footnotesize #2 \end{center}}\end{center}}

\newcommand{\TableSep}[2]{\begin{table}[p]\vspace{#1}
\TabCap{#2}\end{table}}

\newcommand{\FigCap}[1]{\footnotesize\par\noindent Fig.\  %
  \refstepcounter{figure}\thefigure. #1\par}

\newcommand{\TableFont}{\footnotesize}
\newcommand{\TableFontIt}{\ttit}
\newcommand{\SetTableFont}[1]{\renewcommand{\TableFont}{#1}}
\newcommand{\MakeTable}[4]{\begin{table}[htb]\TabCap{#2}{#3}
  \begin{center} \TableFont \begin{tabular}{#1} #4
  \end{tabular}\end{center}\end{table}}

\newcommand{\MakeTableSep}[4]{\begin{table}[p]\TabCap{#2}{#3}
  \begin{center} \TableFont \begin{tabular}{#1} #4
  \end{tabular}\end{center}\end{table}}

\newenvironment{references}%
{ \footnotesize \frenchspacing
\renewcommand{\thesection}{}
\renewcommand{\in}{{\rm in }}
\renewcommand{\AA}{Astron.\ Astrophys.}
\newcommand{\AAS}{Astron.~Astrophys.~Suppl.~Ser.}
\newcommand{\ApJ}{Astrophys.\ J.}
\newcommand{\ApJS}{Astrophys.\ J.~Suppl.~Ser.}
\newcommand{\ApJL}{Astrophys.\ J.~Letters}
\newcommand{\AJ}{Astron.\ J.}
\newcommand{\IBVS}{IBVS}
\newcommand{\PASP}{P.A.S.P.}
\newcommand{\Acta}{Acta Astron.}
\newcommand{\MNRAS}{MNRAS}
\renewcommand{\and}{{\rm and }}
\section{{\rm REFERENCES}}
\sloppy \hyphenpenalty10000
\begin{list}{}{\leftmargin1cm\listparindent-1cm
\itemindent\listparindent\parsep0pt\itemsep0pt}}%
{\end{list}\vspace{2mm}}

\def\TYLDA{~}
\newlength{\DW}
\settowidth{\DW}{0}
\newcommand{\dw}{\hspace{\DW}}

\newcommand{\refitem}[5]{\item[]{#1} #2%
\def\REFARG{#3}\ifx\REFARG\TYLDA\else, {\it#3}\fi
\def\REFARG{#4}\ifx\REFARG\TYLDA\else, {\bf#4}\fi
\def\REFARG{#5}\ifx\REFARG\TYLDA\else, {#5}\fi.}

\newcommand{\Section}[1]{\section{#1}}
\newcommand{\Subsection}[1]{\subsection{#1}}
\newcommand{\Acknow}[1]{\par\vspace{5mm}{\bf Acknowledgements.} #1}
\pagestyle{myheadings}

\newfont{\bb}{ptmbi8t at 12pt}
\newcommand{\xrule}{\rule{0pt}{2.5ex}}
\newcommand{\xxrule}{\rule[-1.8ex]{0pt}{4.5ex}}
\def\thefootnote{\fnsymbol{footnote}}

\begin{center}
{\Large\bf
 The Clusters AgeS Experiment (CASE)  \\
 Analysis of the detached eclipsing binary V15 in 
 the metal-rich open cluster NGC 6253\footnote{Based
 on data obtained with the Magellan, du Pont, and Swope telescopes at Las Campanas Observatory.}}
 \vskip1cm
  {\large
      ~~M.~~R~o~z~y~c~z~k~a$^1$,
      ~~J.~~K~a~l~u~z~n~y$^1$,
      ~~I.~B.~~T~h~o~m~p~s~o~n$^2$,
      ~~A.~~D~o~t~t~e~r$^3$
      ~~W.~~P~y~c~h$^1$
     and  ~~W.~~Narloch$^1$
   }
  \vskip3mm
{ $^1$Nicolaus Copernicus Astronomical Center, ul. Bartycka 18, 00-716 Warsaw, Poland\\
     e-mail: (jka, mnr, wp)@camk.edu.pl\\
  $^2$The Observatories of the Carnegie Institution of Washington, 813 Santa Barbara
      Street, Pasadena, CA 91101, USA\\
     e-mail: ian@obs.carnegiescience.edu\\
  $^3$Research School of Astronomy and Astrophysics, Australian National University, 
      Canberra, Australia\\  e-mail: dotter@mso.anu.edu.au}
\end{center}

\vspace*{7pt}
\Abstract
{We present the first detailed analysis of the detached eclipsing binary V15 in the 
super-metal rich open cluster NGC 6253. We obtain the following absolute parameters: 
$M_p=1.303\pm0.006$~M$_\odot$, $R_p=1.71\pm0.03$ R$_\odot$, $L_p=2.98\pm0.10$ L$_\odot$ 
for the primary, and $M_s=1.225\pm0.006$ M$_\odot$, $R_s=1.44\pm0.02$ R$_\odot$, 
$L_s=2.13\pm0.06$ L$_\odot$ for the secondary. Based on Dartmouth isochrones, the age 
of NGC 6253 is estimated to be 3.80 -- 4.25 Gyr from the mass-radius diagram and 3.9 -- 4.6 
Gyr from color-magnitude diagram (CMD) fitting. Both of these estimates are 
significantly higher than those reported 
so far. The derived apparent distance modulus of 11.65 mag agrees well with the range of 10.9 -- 12.2 
mag derived by other authors; however our estimated reddening (0.113 mag) is lower than the lowest 
published value (0.15 mag). We confirm earlier observations that model atmospheres are 
not accurate enough to account for the whole CMD of the cluster, with the largest 
discrepancies appearing on the subgiant and giant branches. Although age estimation from
the mass-radius diagram is a relatively safe, distance- and reddening-independent
procedure, our results should be verified by photometric and spectroscopic observations
of additional detached eclipsing binaries which we have discovered, at least two of which are 
proper-motion members of NGC 6253.
}
{binaries: close – binaries: spectroscopic – open clusters: 
individual (NGC 6253) – stars: individual (V15-NGC 6253)
}

\Section{Introduction} 
\label{sec:intro}

NGC~6253 is an attractive research target for several reasons.
Because it is an exceptionally metal-rich and evolutionary advanced open 
cluster, it offers a unique opportunity to verify predictions of 
stellar evolution codes in the high-metallicity regime (Claret 2007). 
As a high metallicity environment favors planet formation, it 
is of special interest in the search for exoplanets (Montalto et 
al. 2011) and for the development of the theory of planetary systems. 
Finally, it is an important benchmark for scenarios of the chemical 
evolution of the Galactic disk (Montalto et al. 2011) and an ideal
sample to study the dependence of evolutionary alterations of the chemical
abundances in metal-rich stars (Sestito et al. 2007; Mikolaitis et al. 
2012). 

The cluster has been the subject of several photometric surveys, briefly 
revieved by Kaluzny et al. (2014, hereafter Paper I), a proper-motion
analysis (Montalto et al. 2009), and a radial velocity survey 
(Montalto et al. 2011). Metallicity determinations based on high 
resolution spectra range from ${\rm [Fe/H]}=+0.46$ (Carretta et al.~2007;
Anthony-Twarog et al.~2010) through ${\rm [Fe/H]}=+0.36$ (Sestito et 
al.~2007) to ${\rm [Fe/H]}=+0.19$ (Montalto et al. 2012). The latter 
value is based on observations of just two red giants while the samples
of Carretta et al.~(2007), Anthony-Twarog et al.~(2010) and Sestito et 
al.~(2007) included 4, 15, and 4 stars, respectively. The rms errors
range from 0.03 (Carretta et al.~2007; Anthony-Twarog et al.~2010) to
0.13 (Montalto et al. 2012). Only Carretta et al. (2007) give an estimate 
of the systematic uncertainty (0.08). There is no evidence that 
${\rm [\alpha/Fe]=0.0}$ varies from the solar value. In the present 
paper we adopt ${\rm [Fe/H]}=+0.46$ and ${\rm [\alpha/Fe]=0.0}$.

Age estimates of NGC 6253 have yielded values from 
3.0 Gyr (Bragaglia et al. 1997) to 5.0 Gyr (Piatti et al. 1998). An 
intermediate value of 3.5 Gyr was obtained by Montalto et al. (2009) and 
Sagar et al. (2001). Confidence in all these estimates is lowered by 
significant discrepancies in the photometric results. As we show in Paper I, 
at most one of the four sets of photometry was correctly transformed 
to the standard $BV$ system. Another problem is 
the interstellar reddening, which essentially all authors attempted 
to determine simultaneously with the age of the cluster by means of
isochrone fitting. Additional complications arise from the poorly defined
red giant branch of NGC 6253 and a high contamination of the cluster field by 
background and foreground stars. The astrometric study by Montalto et al. 
(2009) provided a list of likely cluster members; however it also 
demonstrated that on the vector-point diagram the members are poorly 
separated from the field population. 
  
In Paper I, we reported the results of an extensive photometric survey
of NGC~6253 as part of a search for variable stars. Two new detached eclipsing 
binaries were discovered at the turnoff region of the cluster, and another 
one on the subgiant branch. These systems can be used to independently 
determine the age and distance modulus of NGC~6253 as proposed by (Paczy\'nski 
1997). We also obtained good phase coverage for another turnoff binary, 
discovered by Montalto et al. (2011). This object, their star \#45368 and 
our variable V15, is a well-detached SB2 system for which they found 
approximate parameters based on three velocity measurements. Based on 15 
good quality spectra of V15, we report in this paper a substantial refinement 
of the solution obtained by Montalto et al. (2011).  
\section{Observations}
\label{sec:obs}
\subsection{Photometry}
\label{sec:phot}

Our photometric observations of V15 consist of two data sets secured 
at Las Campanas Observatory. $BV$ observations collected with 
the 1.0-m Swope telescope and the SITE3 CCD camera  are described
in detail in Paper I. Additional $BV$ data were obtained
on the 2.5-m du Pont telescope equipped with the SITE2 camera
providing a scale of 0.26 arcsec/pixel. 
Profile photometry was extracted with the Daophot/Allstar utility
(Stetson 1987). Since the field of NGC 6253 is only moderately crowded, the 
quality of the profile photometry was comparable to that obtained with the image 
subtraction technique. Moreover, the profile photometry practically 
eliminated any zero-point differences measured between the 
du Pont and Swope data.
The instrumental magnitudes were transformed to the standard values 
using stars observed by Sagar et al. (2001). A detailed justification for 
this choice of secondary standards is given in Paper I.  

The light curves of V15 turned out to be unstable; most likely due to 
chromospheric activity. Such instabilities are common for short period binaries 
composed of low-mass stars with convective envelopes. We detected variations both in 
and out of the eclipses, with amplitudes reaching 0.03 mag in $V$. Upon averaging 
over seasonal light curves we found $V=14.713\pm0.05$~mag and 
$B-V=0.837\pm0.007$~mag at quadratures.

Altogether, 13 eclipses of V15 were observed between 2007 August 7 and 2013 September 13. 
However, only for a few of these were both ingress and egress covered, which  would allow 
for  precise timing. This prevented us from a classical period study based on 
the determination of individual moments of eclipses and the O-C technique. 
We began by assuming $P=2.572391$~d as found in Paper I and, keeping this value fixed, 
we obtained the photometric solution described in Section \ref{sec:analysis}.
To refine the period, a combined $V$ light curve including data from all seasons 
and both telescopes was used, with seasonal light curves adjusted in magnitude to 
assure a consistent level of the maximum light. The combined curve was fitted with 
the JKTEBOP code\footnote{available at http://www.astro.keele.ac.uk/jkt/codes/jktebop.html}
(Southworth et al. 2004), allowing only for variations of $P$ and the moment of primary 
eclipse $T_{0}$. Mode 8 of the code was used, which enabled a robust determination 
of the errors of the two parameters with the help of a Monte Carlo algorithm. 
We obtained the linear ephemeris 
\begin{eqnarray}
 HJD_\mathrm{min}& = &245 5691.99601(6) + 2.5724149(3)\times E
 \label{eq:ephem}
\end{eqnarray}
where numbers in parentheses are the uncertainties of the last significant 
digits. This refined period was then used to calculate  updated geometric parameters. 
These turned out to be insignificantly different from those reported in 
Section~\ref{sec:analysis}. No evidence for any variability of $P$ was found. The refined 
period is 65.7~s shorter than that found by Montalto  et al. (2011). Their result was based 
on four eclipses, all  only partially covered during the ingress. This difference 
is large enough to cause a profound dephasing of seasonal light curves when their period is used. 

\subsection {Spectroscopy}
\label{subsec:spectra}
Our radial velocity data are based on observations obtained with the 
blue channel of the MIKE Echelle spectrograph (Bernstein et al. 2003) 
on the Magellan Clay telescope between 2010 June 7 and  
2011 July 29 (UT). Most of the observations consisted of two 1200~s exposures 
interlaced with an exposure of a Th/Ar lamp (depending on observing 
conditions, some exposures were shorter or longer). For all observations 
a $0.7\times5.0$ arcsec slit was used, and $2\times 2$ pixel binning 
was applied. For $\lambda=440$~nm the resolution was $\sim$2.7 pixels 
at a scale of 0.0043~nm/pixel.
The typical S/N ratio at the same $\lambda$ was 45. The spectra were 
processed using a pipeline developed by Dan Kelson following the formalism 
of Kelson (2003). 

A total of 15 spectra were used for the analysis. 
The velocities were measured using software based on the TODCOR algorithm 
of Zucker \& Mazeh (1994), kindly made available by Guillermo Torres.
Synthetic echelle-resolution spectra from the library of Coelho et al. 
(2005) with ${\rm [Fe/H]} = +0.46$ and [$\alpha$/Fe]=0.0 were used as velocity 
templates. The templates were Gaussian-smoothed to match the resolution of 
the observed spectra. All velocities were measured on the wavelength range 
400 -- 460 nm. Results of the measurements are presented in Table \ref{tab:vel}. 
\section {Analysis of velocity and light curves}
\label{sec:analysis}
\begin{figure}
   \centerline{\includegraphics[width=0.95\textwidth,
               bb = 20 360 564 690, clip]{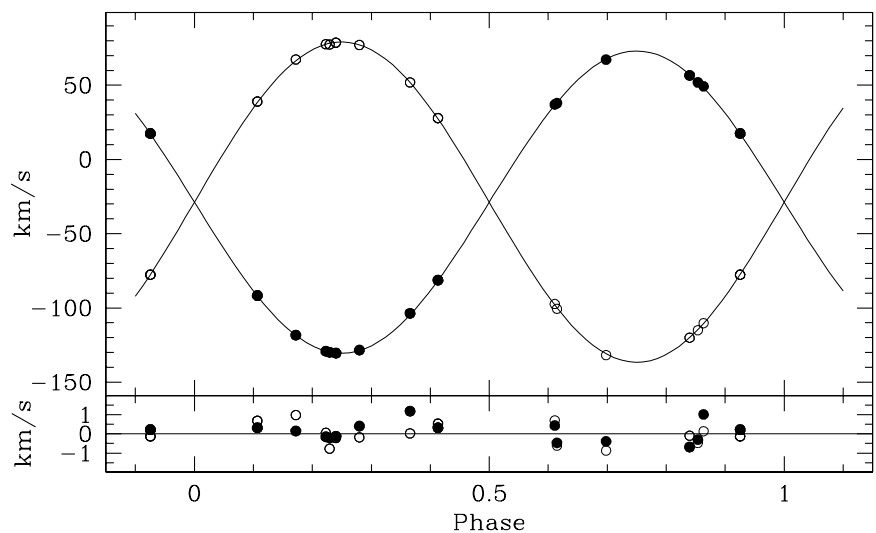}}
   \caption{Velocity curve of V15. Top panel: observed velocities 
    (filled circles: primary; open circles: secondary) and PHOEBE
    fits (lines). Bottom panel: residuals to the orbital fits. 
    \label{fig:vplot}}
\end{figure}

A nonlinear least-squares fit to the observed velocity curves of V15 was 
obtained with the help of  code kindly made available by Guillermo Torres. 
Since the secondary minimum occurs at phase 0.5, the orbital eccentricity 
was fixed at zero while fitting. Observations and orbital solutions
are shown in Fig.~\ref{fig:vplot}, and the derived orbital parameters are 
listed in Table~\ref{tab:orb_parm} together with formal errors returned by 
the fitting routine. The table also lists standard deviations from the orbital 
solution $\sigma_p$ and $\sigma_s$ which are a measure of the precision of 
a single velocity measurement. 
We note that the derived systemic velocity $\gamma$ of V15 agrees well with the mean 
radial velocity of NGC 6253, which according to Montalto et al. (2011) is equal 
to $−29.11\pm0.85)$ km/s.

For the photometric analysis we selected six seasonal curves distinguished by 
low dispersion and relatively high symmetry.
The analysis was performed with the PHOEBE implementation (Pr\v{s}a 
\& Zwitter 2005) of the Wilson-Devinney model (Wilson \& Devinney 1971; Wilson 1979) 
which offers the possibility of simultaneous fitting of $V$ and $B$ light curves. 
Linear limb darkening coefficients were interpolated from the tables of Claret (2000) 
with the help of the JKTLD code.\footnote{Written by John Southworth and available 
at www.astro.keele.ac.uk/jkt/codes/jktld.html}. 

The color of V15 remains nearly constant at all phases, indicating 
nearly equal temperatures of the components. The map of galactic extinction by 
Schlafly \& Finkbeiner (2011) predicts $E(B-V)=0.316$~mag at the location of NGC 
6253. However, the amount of foreground extinction is not known (Anthony-Twarog 
et al. 2010), and values ranging from $E(B-V)=0.23$~mag (Bragaglia et al. 1997) 
to $E(B-V)=0.15$~mag (Montalto et al. 2009) can be found in the literature. Thus 
a direct estimation of the temperature from the dereddened color is not feasible. 
Our procedure, based on Dartmouth isochrones (Dotter et al. 2008) fitted to the 
CMD of the cluster (see Section \ref{sec:discussion}), yields 5830~K and this is 
the value we adopt for both the primary and the secondary, $T_p$ and $T_s$, at 
the start of the PHOEBE iterations.

\begin{figure}
   \centerline{\includegraphics[width=0.95\textwidth,
               bb = 18 408 564 690, clip]{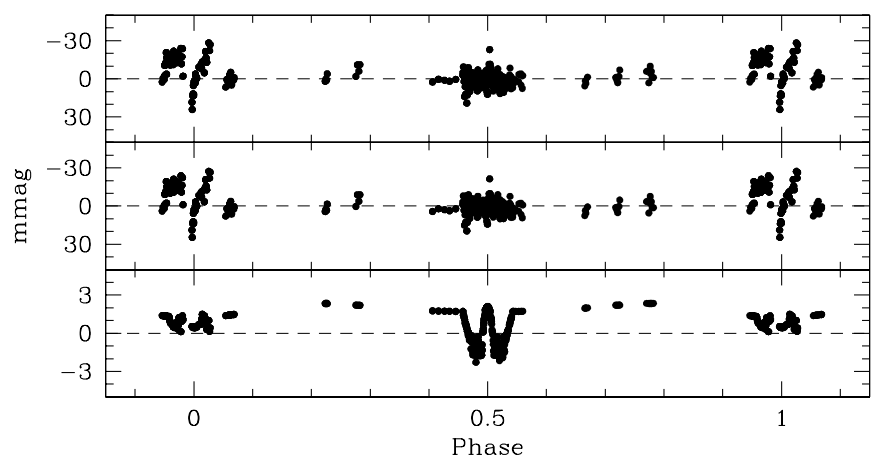}}
   \caption{Illustration of the degeneracy of photometric 
    solutions for V15. From top to bottom: $O-C$ from the $V$ light curve for 
    the solution with $(R_p,R_s)$ = (1.600,1.556) R$_\odot$; the 
    same for the solution with $(R_p,R_s)$ = (1.750,1.406)~R$_\odot$; 
    difference between the two solutions.  
    \label{fig:sequential}}
\end{figure}

\begin{figure}
   \centerline{\includegraphics[width=0.95\textwidth,
               bb = 18 384 564 690, clip]{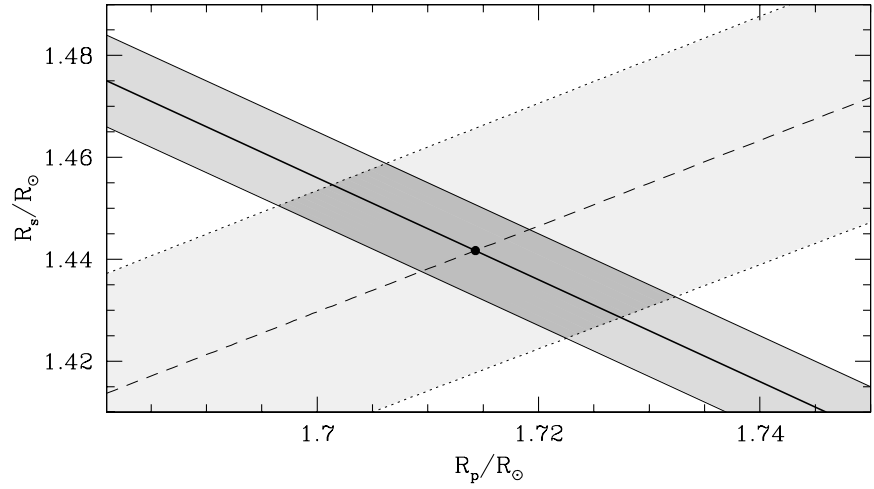}}
   \caption{Photometric fits (grey) and spectroscopic 
            solutions (light-grey) for V15. The black dot marks 
            the best model of the system.
   \label{fig:radii}}
\end{figure}

The eclipses of V15 are partial, moderately deep ($\sim$0.26 mag) and almost equal 
in depth, which causes a well-known degeneracy in the light-curve solutions (Russell \& Merrill 
1952). Keeping $T_p$ fixed and iterating for $T_s$, orbital inclination $i$ and 
surface potentials ($\Omega_p$, $\Omega_s$) we obtained a range of 
equally good fits for which the sum of the component radii was nearly constant at
$3.156\pm0.009$ R$_\odot$. A pair of such fits is shown in Fig. \ref{fig:sequential}. 
Photometric solutions with $3.147<R_p+R_s<3.165$~R$_\odot$ occupy a stripe
signified by the grey colored region in Fig. \ref{fig:radii}. 

\begin{figure}
   \centerline{\includegraphics[width=0.95\textwidth,
               bb = 18 261 564 690, clip]{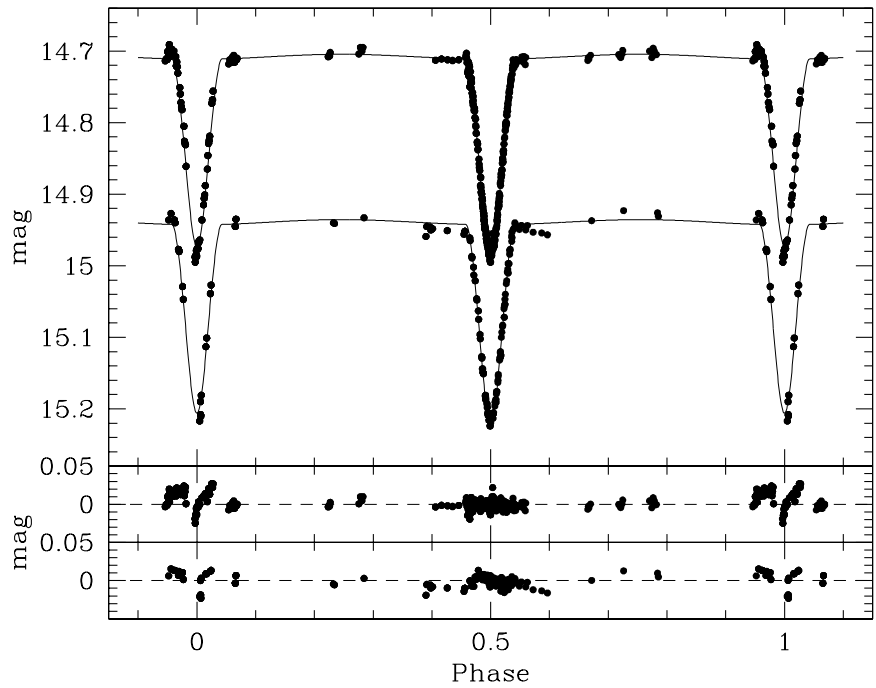}}
   \caption{Top panel: final fits to light curves of V15 (the 
    $B$ light curve is shifted upwards by 0.6 mag). Middle panel: 
    $V$ residuals. Bottom panel: $B$ residuals. 
    \label{fig:lcplot}}
\end{figure}

To remove the degeneracy we calculated a series of synthetic spectra of V15, again 
using the library of Coelho et al. (2005). The spectra retrieved from the library were
rotationally broadened (a synchronous rotation of both components was assumed) and 
Doppler-shifted to the velocities of the primary and secondary listed in Table \ref{tab:vel}. 
These pairs of spectra corresponding to a given phase were then combined 
in various proportions and compared to the observed spectrum taken at the same phase. 
The comparison was performed separately for 15 different spectral ranges between 
408 and 496~nm, each 3 nm long (we decided to compare short segments of the spectra 
rather than the whole available range in order to account for the varying mean intensity).
For each phase and each segment the best value of the total secondary-to-primary light ratio 
was found by minimizing the sum of squared differences in synthetic and observed 
spectrum intensity. For further analysis the mean value of light ratios thus obtained, 
$q_l=0.707\pm0.024$, was used. 
From $q_l$ one can derive the ratio of the radii of the components 
of the binary  $$R_s/R_p=\sqrt{q_l\beta(T_p)/\beta(T_s)},$$ where 
$$\beta(T)=\int B_\nu(T) s(\nu) d\nu,$$ $B_\nu(T)$ is the Planck function and $s(\nu)$ 
is the spectral sensitivity of MIKE's blue arm. Since the difference between $T_s$ 
and $T_p$ is small, we can write $$R_s/R_p=\sqrt{q_l}(1-2(T_s-T_p)/T_s).$$ 
We found that the correction introduced by the temperature factor 
was smaller than 0.005 and could be neglected. Thus, the spectroscopic data impose a 
condition $R_s/R_p=0.841\pm0.014$ which on the $(R_p,R_s)$ plane defines a stripe 
marked with the light grey 
color in Fig.~\ref{fig:radii}. The best solution is defined by the intersection 
of lines $R_p+R_s=3.156$ R$_\odot$ and $R_s/R_p=0.841$. At the intersection we 
have $R_p=1.714 \pm0.018$ R$_\odot$ and $R_s=1.441\pm0.018$ $R_\odot$, where the 
errors are defined by the corners of the dark-grey quadrangle in Fig.~\ref{fig:radii}.
The remaining parameters of the best photometric model of V15 can be found in 
Table~\ref{tab:phot_parm}, and the final fits to the observed light curves are 
shown in Fig.~\ref{fig:lcplot}. The errors of $i$, $T_s$, $(L_p/L_s)_V$ and 
$(L_p/L_s)_B$ given in Table~\ref{tab:phot_parm} are based on 20,000 Monte Carlo 
simulations performed with the help of a procedure written in PHOEBE-scripter after 
that outlined in the description of the JKTEBOP code (Southworth et al. 2004, 
and references therein). Upon combining data from Tables \ref{tab:orb_parm} and 
\ref{tab:phot_parm} we obtained the absolute parameters of V15 listed in 
Table~\ref{tab:abs_parm}. 

We finish this Section with a word of caution. One should be very careful when 
dealing with light curves similar to that of V15, as accepting the first solution 
found may lead one completely astray. The problem becomes especially acute when 
automatic light-curve solvers are applied. 

\Section{Discussion}
 \label{sec:discussion}

\begin{figure}
   \centerline{\includegraphics[width=0.95\textwidth,
               bb = 54 266 564 690, clip]{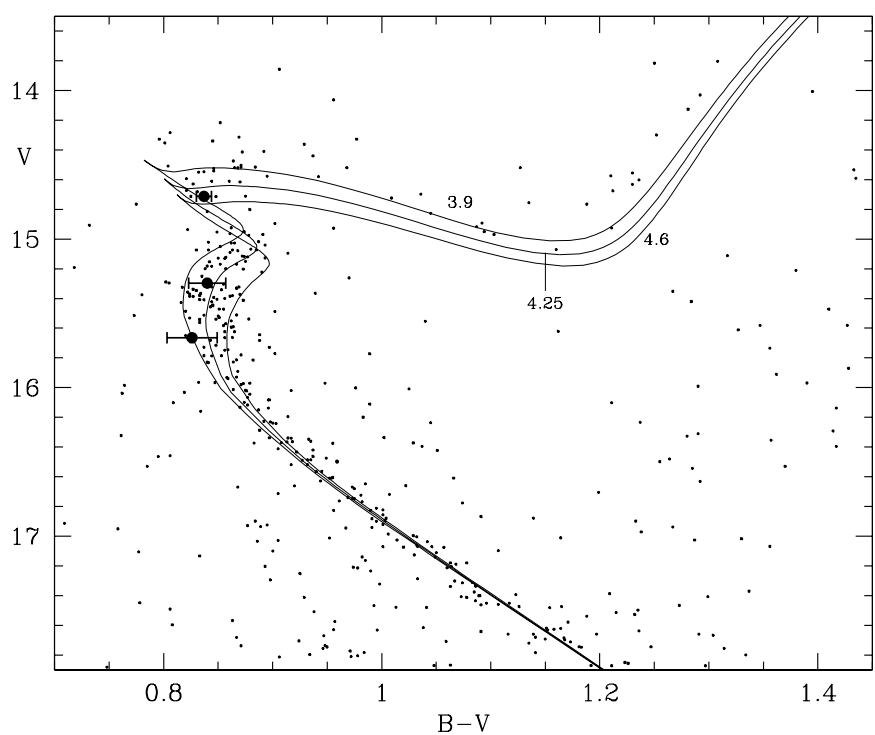}}
   \caption{The color-magnitude diagram of NGC 6253 with fitted Dartmouth 
    isochrones. The apparent distance modulus and $E(B-V)$ resulting from 
    the fit are equal to 11.65 mag and 0.113 mag, respectively. Labels 
    indicate ages in Gyr. Large dots show locations of V15 and its 
    components (errors in $V$ are too small to be visualized on this 
    scale). Only proper motion and radial velocity members of the cluster 
    are plotted, following Montalto et al. (2009; 2011). 
    \label{fig:cmd}}
\end{figure}

Fig.~\ref{fig:cmd} shows the color-magnitude diagram of NGC 6253 to which Dartmouth 
isochrones with [Fe/H] = +0.46 and [$\alpha$/Fe]=0.0 are fitted. 
Simultaneous fitting of unevolved main sequence, turnoff region, subgiant branch 
and giant branch turned out to be impossible. Similar problems had been encountered 
earlier by Anthony-Twarog et al. (2007), Montalto et al. (2009), and Anthony-Twarog 
et al. (2010). Since uncertainties of stellar models 
grow larger with evolutionary time, we decided to assign the largest weight to main 
sequence and turnoff. We found the latter to be 
nearly entirely contained between isochrones for 3.9 and 4.6 Gyr, with that for 4.25 Gyr 
running through its center (Fig.~\ref{fig:cmd}). The mean turnoff temperature obtained 
from the three curves amounts to 5830~K with the caveat that the fit yields 
$E(B-V)=0.113$~mag, i.e. a value lower than the lowest used so far. Keeping this in mind, 
we performed a consistency check using dereddended color index of the cluster and 
color-temperature calibration of Sousa et al. (2011) which is valid for 
$-1.5<$[Fe/H/]$<+0.5$, $0.4<B-V<1.2$ mag and $4500<T<6400$~K. Applying 
this calibration to 
[Fe/H] = +0.46 and $(B-V)_0=0.724$ mag yields $T=5797$ K -- a value consistent with 
the assumed one within the 1-$\sigma$ range of the calibration, equal to 52 K. 

Fig.~\ref{fig:cmd} also shows the location of V15 and its components on the CMD of
the cluster. Both the primary and the secondary are located at the turnoff, 
with the primary being slightly more evolved.
The activity of V15 suggests that its components might be cooler and larger than 
inactive stars of the same mass and at the same evolutionary phase (see e.g. Morales 
et al. 2008). However, there are good reasons to believe that they are quite normal.
First, none of our spectra shows emission in Ca II H \& K or the Balmer lines. 
Second, both components are located close to the 
blue edge of the turnoff, whereas one would expect them to be shifted to the red 
if their sizes and temperatures are affected by the activity.

The apparent distance modulus obtained from the isochrone fit is 11.65 mag. This is 
nearly the mean of values found by the other authors, which range from 
10.9 mag (Sestito et al. 2007) to 12.2 mag (Twarog et al. 2003). 

\begin{figure}
   \centerline{\includegraphics[width=0.95\textwidth,
               bb = 32 323 565 690, clip]{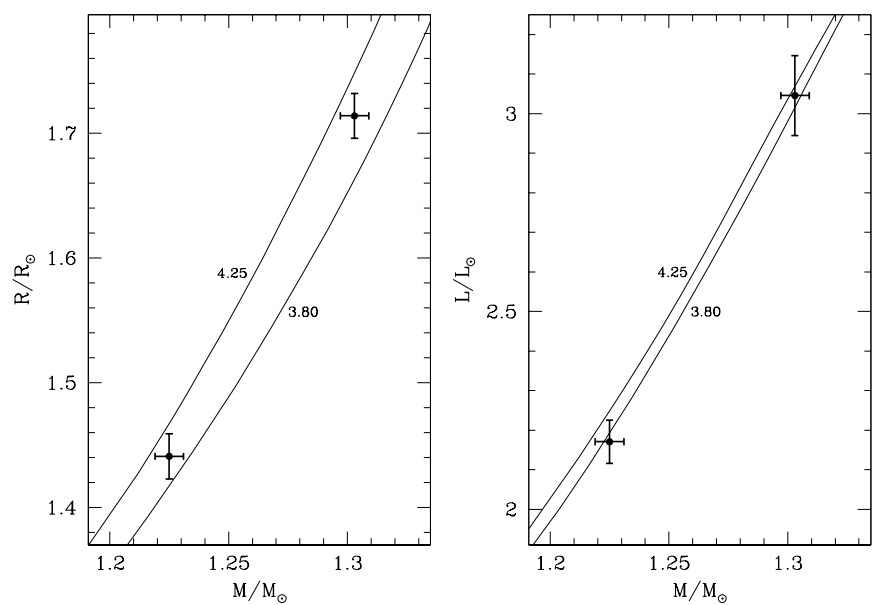}}
   \caption{Dartmouth isochrones for [Fe/H] = +0.46, 
    [$\alpha$/Fe]=0.0 compared to values found for the components of V15  
    in the mass-radius diagram (left) and mass-luminosity
    diagram. Labels indicate ages in Gyr. 
    \label{fig:age}}
\end{figure}

Almost all ages reported so far for NGC 6253 range from 2.5 Gyr 
(Anthony-Twarog et al. 2007) to 3.5 Gyr (Montalto et al. 2009), the 
sole exception is an age of 5 Gyr found by Piatti et al. (1998). Our attempts 
to fit Dartmouth isochrones for those ages failed completely. For the younger 
ages the discrepancy between theoretical turnoff temperature and temperature 
obtained from the calibration was too large, while for the older ages the 
reddening became unreasonably small ($<0.08$ mag). Further support for our 
CMD fit comes from the mass-radius diagram shown in Fig. \ref{fig:age},
where error boxes of both components are contained between isochrones for 
3.8 Gyr and 4.25 Gyr. As the mass-radius relation is free from uncertainties 
in distance and extinction that plague ages derived via isochrone fitting, 
we regard this estimate as very reliable. The data are also compatible with 3.8 Gyr and 4.25 Gyr
isochrones in the mass-luminosity diagram (Fig. \ref{fig:age}) , 
but because of the way the temperature was estimated this result 
cannot be regarded as an independent verification.    

The recently published BT-Settl model atmospheres (Allard 2014)  
produce bluer isochrones than those from the Dartmouth database. We 
obtained several such isochrones and repeated CMD and mass-radius fitting. 
The age derived from the mass-radius diagram did not change; essentially 
the same age as before (3.8-4.5 Gyr) was also derived from the CMD fit. 
However, the distance modulus and reddening changed to 11.45 mag and 0.13 mag, 
respectively. The overall agreement of isochrones with the data did not 
improve. While most objects on the subgiant branch fell within the area 
bordered by 3.8 Gyr and 4.5 Gyr lines (as opposed to Fig. \ref{fig:cmd},
where most of them lie above the 3.9 Gyr line), all BT-Settl isochrones
missed the "blue clump" composed of seven stars surrounding the location 
of the binary. The O-C discrepancy on the giant branch remained as large
as before, however the theoretical colors were bluer 
than the observed ones. 

Brogaard et al. (2012) extensively discuss problems related to the 
theoretical relation between colours and effective temperatures, which 
are a major uncertainty when comparing observations to stellar models 
in the CMD. The problem becomes especially acute at high metallicities, 
where a significant contribution to opacity comes from metals with poorly 
known abundances. Brogaard 
et al. (2012) explicitly warn that a good match between models and the 
observed CMD is likely to be more of a coincidence than a reflection 
of reality. Obviously, 
the same may be true regarding age estimation based on CMD fitting. 
Indeed, the agreement between ages derived for V15 from CMD and 
mass-radius diagram is so good that a suspicion arises it might be 
spurious. Although age determination from the mass-radius diagram is 
a relatively safe procedure, it is clear that our results should be 
verified. 

In Paper I we reported a discovery of additional three detached eclipsing binaries 
in NGC 6253, two of which are proper motion members of the cluster (PM data are 
missing for the third one). All these systems are bright enough 
($V_{\mathrm max}<15.2$ mag) for detailed spectroscopic studies. Solving for 
their light and velocity curves would create a very valuable benchmark against 
which stellar evolution codes could be tested in the high metallicity regime 
through mass-radius and mass-luminosity comparisons. Furthermore, a significant 
improvement in CMD fitting is possible provided that better proper motion and 
radial velocity data will be collected (the presently available ones are not 
sufficient to select cluster members with confidence; see Fig. \ref{fig:cmd}). 
Future observations should also determine the reddening in a way independent of
isochrone fitting. Anthony-Twarog et al. (2010) discuss this issue in detail,
and show that the task is by no means easy. 

\Section{Summary}
 \label{sec:sum}
Based on photometric and spectroscopic observations of the detached eclipsing 
binary V15 in NGC 6253 we derived absolute parameters of its components. Both 
the primary and the secondary are located at the turnoff of the cluster,
making them suitable for an age estimate from the mass-radius diagram. Using 
Dartmouth isochrones we find NGC 6253 to be 3.80 -- 4.25 Gyr old - a range 
of ages compatible with 3.9 -- 4.6 Gyr derived from CMD fitting. Both these 
estimates are significantly higher than those reported so far, which with 
one exception do not exceed 3.5 Gyr. The apparent distance modulus found 
from CMD fitting amounts to 11.65 mag and it agrees well with 10.9 -- 12.2
mag derived by other authors; however the reddening (0.113 mag) is lower 
than the lowest published value (0.15 mag). We confirm earlier conclusions 
that models of metal-rich atmospheres are not accurate enough to account for 
the whole CMD of the cluster, with the largest discrepancies appearing at 
advanced evolutionary phases. Although age estimation from the mass-radius 
diagram is a relatively safe, distance- and reddening-independent procedure, 
we stress the need to verify our results by photometric and spectroscopic 
observations of three detached eclipsing binaries discovered by Kaluzny 
et al. (2014), at least two of which are proper-motion members of NGC 6253.

\Acknow
{
JK, WN, WP and MR were partly supported by the grant DEC-2012/05/B/ST9/03931
from the Polish National Science Center. 
}

\clearpage
\begin{table}
\caption{Radial velocity observations of V15 
\label{tab:vel} 
     }
 \centering
 \begin{tabular}{@{}lrrrrr}
 \hline
 \hline
HJD      & $v_{p}$   &  $v_{s}$ & Phase  & $(O-C)_{p}$ & $(O-C)_{s}$\\
-2450000 & [km/s]    &  [km/s]  &        & [km/s]      & [km/s] \\
\hline
5354.59687	&	56.61	&	-120.07	&	0.840	&	-0.72	&	-0.15\\
5354.81742	&	17.45	&	-77.63	&	0.925	&	0.18	&	-0.18\\
5355.62665	&	-130.44	&	78.70	&	0.240	&	-0.18	&	-0.25\\
5388.72531	&	-91.68	&	38.97	&	0.107	&	0.27	&	0.64\\
5389.51319	&	-81.33	&	27.85	&	0.413	&	0.27	&	0.49\\
5457.55448	&	49.27	&	-110.26	&	0.863	&	0.97	&	0.09\\
5458.47928	&	-129.19	&	77.62	&	0.223	&	-0.20	&	0.02\\
5459.48720	&	37.94	&	-100.56	&	0.615	&	-0.50	&	-0.66\\
5763.64679	&	51.86	&	-115.00	&	0.854	&	-0.34	&	-0.52\\
5764.61295	&	-129.86	&	77.45	&	0.229	&	-0.25	&	-0.80\\
5770.73991	&	37.08	&	-97.38	&	0.611	&	0.39	&	0.66\\
5836.49309	&	-118.38	&	67.40	&	0.172	&	0.10	&	0.94\\
6096.58400	&	-128.34	&	77.07	&	0.280	&	0.36	&	-0.22\\
6096.80512	&	-103.62	&	51.89	&	0.366	&	1.13	&	-0.01\\
6097.66008	&	67.25	&	-131.80	&	0.698	&	-0.42	&	-0.92\\
\hline
\end{tabular}
\end{table}

\begin{table}
 \caption{Orbital parameters of V15
          \label{tab:orb_parm}
         }
 \centering
 \begin{tabular}{lcc}
  \hline
   Parameter        & Value &Error   \\
  \hline
  $\gamma$ (km s$^{-1}$)     & -28.73 &0.011  \\
     $K_p$ (km s$^{-1}$)     & 101.61 &0.20  \\
     $K_s$ (km s$^{-1}$)     & 108.06 &0.19  \\
     $q$                     & 0.9403  &0.0027 \\
     $e$                       & 0.0$^{\mathrm a}$& \\
     $\sigma_p$ (km s$^{-1}$)   & 0.51 &           \\
     $\sigma_s$ (km s$^{-1}$)   & 0.54 &             \\
     Derived quantities:        &       &          \\
     $A\sin i$ (R$_\odot$)     & 10.657 &0.015  \\
     $M_p\sin^3 i$ (M$_\odot$) & 1.2662 &0.0055   \\
     $M_s\sin^3 i$ (M$_\odot$) & 1.1906 &0.0056  \\
\hline
 \end{tabular}\\
\rule{0 mm}{3 mm}
$^\mathrm{a}$Assumed in fit\\
\end{table}

\begin{table}
 \caption{Photometric parameters of V15
          \label{tab:phot_parm}
         }
\centering
 \begin{tabular}{lcc}
  \hline
   Parameter        & Value & Error\\
  \hline
     $i$(deg)       &   82.14           & 0.06  \\
     $R_p$          &   1.714           & 0.018   \\
     $R_s$          &   1.441           & 0.018   \\
     $e$            & 0$^\mathrm{a}$    &   \\
     $T_{p}$ (K)    & 5830$^\mathrm{a}$ &   \\
     $T_{s}$ (K)    & 5842              & 10\\
     $(L_{p}/L_{s})_{V}$  & 1.403       & 0.033   \\
     $(L_{p}/L_{s})_{B}$  & 1.393       & 0.034   \\
     $\sigma_{rms}(V)$ (mmag) & 7   &     \\
     $\sigma_{rms}(B)$ (mmag) & 7  &    \\
     $V_p$ (mag)& 15.297  & 0.012$^\mathrm{b}$ \\
     $V_s$ (mag)& 15.665  & 0.016$^\mathrm{b}$ \\ 
     $B_p$ (mag)& 16.137  & 0.012$^\mathrm{b}$ \\
     $B_s$ (mag)& 16.497  & 0.016$^\mathrm{b}$ \\
     $(B-V)_{p}$(mag)& 0.840& 0.017$^\mathrm{b}$ \\ 
     $(B-V)_{s}$(mag)& 0.832& 0.023$^\mathrm{b}$ \\ 
  \hline
 \end{tabular}\\
\rule{0 mm}{3 mm}$^\mathrm{a}$Assumed.\\
\rule{1 mm}{0 mm}$^\mathrm{b}$Includes errors from photometric solution and 
profile photometry.
\end{table}
\begin{table}
 \caption{The physical properties of V15
          \label{tab:abs_parm}
         }
 \centering
 \begin{tabular}{lcc}
  \hline
    Parameter        & Value &Error   \\
  \hline
     $M_p$ (M$_\odot$) &1.303& 0.006 \\
     $M_s$ (M$_\odot$) &1.225& 0.006 \\
     $R_p$ (R$_\odot$) & 1.714& 0.018 \\
     $R_s$ (R$_\odot$) & 1.441& 0.018 \\
     $T_p$ (K) & 5830$^\mathrm{a}$ &   \\
     $T_s$ (K) & 5842&10  \\
     $L^{bol}_{p} $ (L$_\odot$) &3.05& 0.10 \\
     $L^{bol}_{s} $ (L$_\odot$) &2.17& 0.06 \\
     $A$ (R$_\odot$) & 10.758 & 0.017\\
     $P$ (d)& 2.5724149& 3$\times10^{-7}$\\
\hline
 \end{tabular}\\
\rule{0 mm}{3 mm}$^\mathrm{a}$Assumed.\\
\end{table}

\clearpage

\end{document}